\documentclass{aa}  
\usepackage{graphicx}
\usepackage{txfonts}
\usepackage{natbib}
\usepackage{epsfig}
\bibpunct{(}{)}{;}{a}{}{,}
\def\micron{$\mu$m}

\begin{document}

\title{A narrow-band search for Ly$\alpha$ emitting galaxies at
$z=8.8$\thanks{Based on observations collected at the European Southern
  Observatory, Chile, Programs 072.A-0454, 076.A-0775.}}

   \author{J.-G. Cuby 
          \inst{1}
          \and
          P. Hibon\inst{1,2}
          \and
          C. Lidman\inst{2}
          \and
          O. Le F\`evre\inst{1}
          \and
          R. Gilmozzi\inst{3}
          \and
          A. Moorwood\inst{3}
          \and
          P. van~der~Werf\inst{4}
          }

\institute{Laboratoire d'Astrophysique de Marseille, B.P. 8, F-13376 Marseille
           Cedex 12, France 
           \and
           European Southern Observatory, Alonso de Cordova 3107, Vitacura,
           Casilla 19001, Santiago 19, Chile
           \and
           European Southern Observatory, Karl-Schwarzschild-Str. 2,
                D-85748 Garching, Germany
           \and 
           Leiden Observatory, P.O. Box 9513, NL-2300 RA Leiden, The
           Netherlands
          }

\date{Recieved; accepted}


\abstract
{}
{The first star forming galaxies in the early universe should be copious
Ly$\alpha$ emitters, and may play a significant role in ionizing the
intergalactic medium (IGM). It has been proposed that the luminosity
function of Lya emitting galaxies beyond z$\sim$6 may be used to constrain the
neutral fraction of the IGM during this epoch. In this work we report on a
search for Ly$\alpha$ emitters at redshift 8.8.}
{We performed a narrow band imaging programme using ISAAC at the ESO VLT.
Seven fields, covering a total area of 31
sq. arcmin and for which optical and broad band infra-red images have
been obtained in the GOODS survey, were imaged to a limiting flux
(respectively luminosity) of
$\sim 1.3 \times 10^{-17}$\,ergs\,s$^{-1}$\,cm$^{-2}$
(respectively $\sim 1.3 \times 10^{43}$\,ergs\,s$^{-1}$)
in a narrow band filter centered
in a region of low OH sky emission at 1.19\,\micron. Candidate Lyman$\alpha$
emitters are objects that are detected in the ISAAC NB images
and undetected in the visible broad band images.}
{No $z=8.8$ Ly$\alpha$ emitting
galaxies were detected to a limit approaching recent estimates of the
luminosity function at $z\sim 6$.  Our results do suggest, however, that
detections or substantial constraints could be achieved by this method
in the near future with larger field instruments planned for various
telescopes.}
{}


   \keywords{Galaxies: luminosity function -- (Cosmology): early Univserse}

   \maketitle


\section{Introduction}

The reionisation of the universe is intricately linked to the formation
of the first objects. While observations of high redshift QSOs
indicate that the end of reionisation occurred around a redshift of 6
\citep{Fan2002}, the measurement of the optical depth to electron
scattering in the Wilkinson Microwave Anisotropy Probe (WMAP)
three year results indicates that the universe
was partially ionised up to significantly higher redshifts
\citep{Page2006,Spergel2006}. Hence, the current data
point to an extended period of reionisation, ending at a redshift of 6
and extending to redshifts well above 10.

The objects that cause reionisation need not be the same throughout
this period.  A plausible model that matches the WMAP three year data
is a model in which Population II and Population III objects
contribute equally to reionisation \citep{Wyithe2006}. In this model,
the ionised fraction stays approximately constant over the redshift
interval $z=7-12$. However, other models, leading to quite different
ionised fractions in this redshift interval, are also possible, so
observational constraints on the fraction of ionised gas are now quite
important.

The first star forming objects in the universe should be copious
emitters in Ly$\alpha$ and may play a significant role in ionising the
intergalactic medium (IGM). It has been proposed that the luminosity
function of Ly$\alpha$ emitters (LAEs) may be used to constrain the
neutral fraction of the IGM when the neutral fraction is as high as
30\% \citep{Malhotra2004}, since the red wing of the Gunn-Peterson
trough should attenuate the Ly$\alpha$ line. As a comparison, the
Gunn-Peterson trough measured blueward of the Ly$\alpha$ line
saturates when the neutral fraction reaches 1\% \citep{Fan2002}.

Current results on the luminosity function of LAE at $z=6.5$ are
contradictory. While \citet{Malhotra2004} and \citet{Taniguchi2005} find
no evidence for evolution in the LAE luminosity function between $z=5.7$
and and $z=6.5$, thus suggesting that the universe was still largely
reionised at $z=6.5$, \citet{Kashikawa2006} find significant evolution.



Narrow-band, wide-field imaging surveys are an efficient and
effective method of discovering LAEs. The narrow band filters are
designed to cover regions where the night sky
background is low. This naturally leads to surveys occurring at
discrete redshifts, such as $z=3.4$, corresponding to 5390\,\AA\
\citep{Cowie1998}; $z=4.5$, corresponding to 6740\,\AA\
\citep{Hu1998,Malhotra2002}; z=$5.7$, corresponding to 8150\,\AA\
\citep{Rhoads2003,Ajiki2006,Hu2004,Shimasaku2006,Westra2006} and
$z=6.5$, corresponding to 9200\,\AA\
\citep{Hu2002a,Hu2002b,Kodaira2003,Cuby2003,Taniguchi2005}.  

To detect Lyman alpha emitters (LAE) at still higher redshifts, one
has to move into the near IR. The next two windows that are relatively
free of bright OH lines occur at 1.06\,\micron\ and 1.19\,\micron, and
correspond to LAEs at $z=7.7$ and $z=8.8$, respectively. In a
pioneering study, \citet{Willis2005} used ISAAC on Antu (VLT-UT1) to
image a single field in the Hubble Deep Field South with a narrow band
filter centred 1.19 \micron. In a volume of $340\,h^{-3}\,{\rm Mpc}^3$
they find no LAEs brighter than the survey limit of 
$10^{43}\,h^{-2}\,{\rm ergs\,s}^{-1}$.

In this paper, we complement the work of \citet{Willis2005} by using
the same instrument and filter to image a wider region of the sky to a
brighter survey limit. In sections 2 and 3, we describe the new narrow
band IR imaging data obtained in this survey and how we combine these
data with broad band optical and IR data from the GOODS survey to
search for LAEs at $z=8.8$. In section 4, we describe spectroscopic
follow-up of our most promising candidate, and in section 5 we
estimate the likelihood of detecting supernovae and solar system
objects as false candidates. In section 6, we place limits on the
$z=8.8$ luminosity function, and we discuss the
possibility of extending this study over a wider area. Throughout this
paper we use a flat $\Lambda$CDM model with $\Omega_M=0.3$ and
$H_0=70\,{\rm km\,s}^{-1}$ to compute distances and volumes.

\section{ISAAC observations and data reduction}

\subsection{ISAAC narrow band observations}

We used the short wavelength (SW) arm of ISAAC \citep{Moorwood1998} on
Antu (VLT-UT1) to image selected fields with the ISAAC narrow band
filter centred at 1.19 micron. The detector in the SW arm of ISAAC is
a Hawaii HgCdTe array, the pixel scale is 0\farcs148 and the field-of-view is
2\farcm5 by 2\farcm5.

All the data were taken in service mode over a
period lasting several months. The observations started on 2004
November 13 and finished on 2005 March 21 and occurred on 20 different
nights.

To ensure background limited performance, each exposure lasted
approximately 20 minutes. After each exposure, the telescope was
offset by approximately 20\arcsec\ and a new exposure taken. This
sequence was repeated while conditions were good enough and typically
3 to 6 exposures were taken on any given night. 

We targeted seven ISAAC fields within the Chandra Deep Field South
(CDF-S) field covered by the Great Observatories Origins Deep Survey
(GOODS). A list of the fields and the corresponding exposure times and
limiting magnitudes are given in Table \ref{tab_gen}. The total area
of the survey (defined as being the sum of the areas that were
continuously imaged) was 31 square arc minutes.  This area has been
extensively studied at multiple wavelengths - from the X-ray to the
mid-IR. For the purposes of this study we make use of publicly
available near-IR (VLT/ISAAC) (version 1.5) and optical (HST/ACS) data
(version 1.0).

\begin{table}[htbp]

\caption{The fields observed and their coordinates, the exposure
times, the magnitude at which the narrow band data are 50\% complete
and the image quality of the stacked images. All magnitudes are in
Vega and have been corrected to 4\arcsec\ diameter apertures. The 90\%
completeness limit is approximately 0.5 magnitude brighter. The field
names correspond to those used at http://www.eso.org/science/goods/}


\label{tab_gen}

\centering
\begin{tabular}{cccccc} 
\hline\hline
Field  & RA & DEC & Exp.  & Lim.  & IQ  \\
&(2000)&(2000)& time            & mag.    & FWHM    \\
&&&  (s)          &  (Vega)   & (\arcsec)     \\
\hline
F09    & 03:32:33.2    & -27:44:55    & 10800         & 23.4 & 0.55   \\
F11    & 03:32:10.6    & -27:44:55    & 18000         & 23.2 & 0.46   \\
F14    & 03:32:38.8    & -27:47:25    & 21600         & 23.2 & 0.55   \\
F15    & 03:32:27.5    & -27:47:25    &  6000         & 22.4 & 0.69   \\
F16    & 03:32:16.2    & -27:47:25    & 26400         & 23.2 & 0.53  \\
F20    & 03:32:33.2    & -27:49:55    & 18000         & 23.3 & 0.59   \\
F21    & 03:32:21.9    & -27:49:55    & 20400         & 23.3 & 0.58  \\
\hline

\end{tabular}
\end{table}

\subsection{ISAAC narrow band reduction}

Apart from the removal of the background, the methods used to reduce
the ISAAC narrow band data were standard. A dark frame was first
subtracted and a twilight flatfield applied. The removal of the
background was non-standard and had to be done in two
steps. Usually, when processing data in the near IR, the background of
any particular frame is estimated by averaging the frames that are
taken immediately before and after the frame in question. To maximize
the signal-to-noise ratio, one should use as many frames as possible,
but this is limited either by the number of frames that are available
or by the timescale over which the background varies. In these data, 
the large scale component of the background was varying with a
timescale that was similar to the length of the exposure. 

The reason for the variation is that, within the field of view of ISAAC,
the central wavelength of the narrow-band filter transmission curve
varies. In the upper part of the field, light from OH lines pass
through the shifted filter band pass. Since the intensity of OH lines
can vary considerably over a period of a few minutes and since
exposures were relatively long (20 minutes), there was considerable
variation in the large scale background, especially in the upper
part of the array.

In addition to the relative rapid changes in the background,
sometimes, there were as few as three frames taken consecutively, so,
in principle, one had only two frames to use as sky.

In order to remove the background adequately and to maximize the
signal-to-noise ratio in the background subtracted images, we first
removed the large scale shape of the background in individual frames
by applying a median filter with a box size of 200 pixels to each
image and by subtracting the result. Once the large scale shape is
removed, we then collect all the data for a given field, even if it
was taken on different nights, and follow the standard procedure of
background subtraction that is descibed above. This unusual approach only
works if the small scale variations (including pixel-to-pixel
variations) are stable over the time the data are taken. This appears to
be the case for these data as the pixel-to-pixel noise in the reduced
data is within a few percent of the theoretical expectation.

We used the XDIMSUM package in IRAF\footnote{IRAF is distributed by
the National Optical Astronomy Observatories, which are operated by
the Association of Universities for Research in Astronomy, Inc., under
the cooperative agreement with the National Science Foundation.} to do
the background subtraction as it allows object and bad pixel masking
when computing the background frames. Typically, a background frame is
made up of 15 to 20 frames.

The alternative of subtracting the backgound first and then removing
any large scale residuals with a median filter proved to be
unsatisfactory, as objects, variable pixels and cosmic rays cannot be
properly identified by the XDIMSUM software when the background
variation between frames is as large as it was in these data.

The background subtracted images were then corrected for field
distortion, using the field distortion coefficients available from the
ISAAC web pages\footnote{http://www.eso.org/instruments/isaac/},
registered, combined and astrometrically aligned to the broad band
ISAAC and ACS images. Because some of these steps involve
interpolation, the pixel-to-pixel noise in the fully reduced images is
a few percent less than the value one would derive from the mean level
of the background.

\section{Analysis}

\subsection{Catalogue generation and detection limits}

We used the SExtractor (Bertin \& Arnouts,1996) software in single
image mode to detect and measure objects in the ISAAC narrow band
images and in the ISAAC and ACS broad band images. We created one
catalogue per filter per field and we then used our own software to
match the catalogues in sky coordinates. The matching radius was set
to between 0.2 and 0.4 arc seconds, depending on the field.

The magnitudes were first computed for apertures of 8 pixels in
diameter and were then corrected to apertures of 4\arcsec\ in diameter
by applying an aperture correction that was computed from bright
unsaturated stars.

\begin{figure}

   \resizebox{\hsize}{!}{\includegraphics[width=8cm,height=12cm]{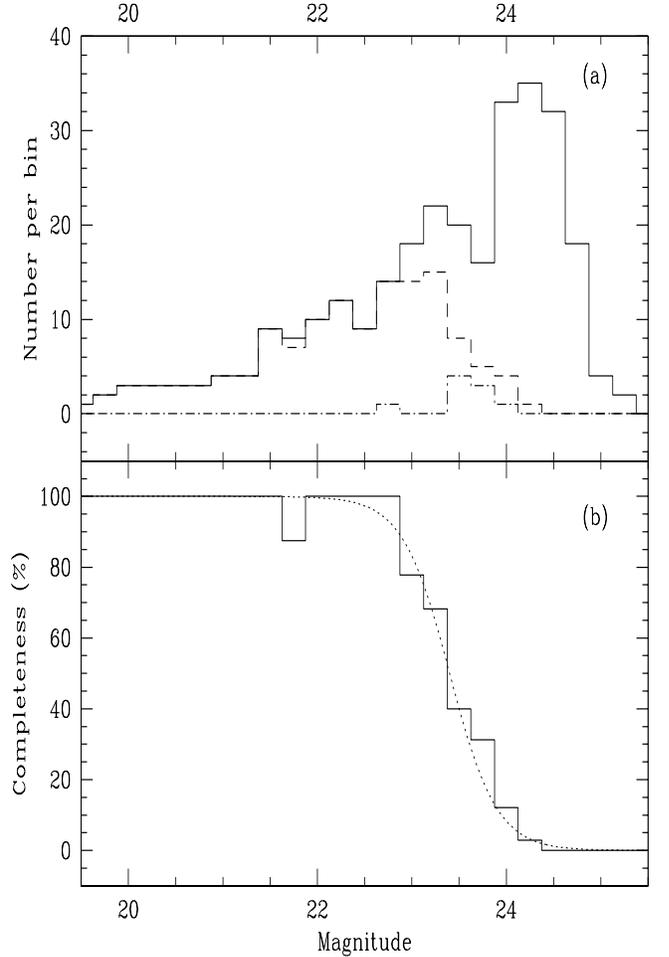}}

   \caption{{\bf Panel (a):} A comparison of the number of objects
   detected in the J-band image (solid line), in both the narrow band
   and J band images (dashed line), and in the narrow band image alone
   (dot-dashed line). {\bf Panel (b):} A histogram of the
   completeness function of objects detected in the narrow band
   filter. The histogram corresponds to the ratio between the dashed and
   solid lines in the upper panel. The dotted line is a fit
   to the histogram using equation \ref{eq:completeness}.}

   \label{fig:incompleteness}

\end{figure}

The limiting magnitude of the narrow band data can be estimated by
comparing the number of detections in the narrow band image with the
number of detections in the much deeper ISAAC J-band image. Since the
ISAAC image is much deeper and since ${\rm J-NB1190} \sim 0$ for most
objects, objects detected in the narrow band filter should also be
detected in J.  By plotting the ratio of the number of objects seen in
both the narrow band image and the J-band image over the number of
objects seen in the J-band image, one can get an estimate of the
completeness. An example for one of the fields is shown in
Fig.~\ref{fig:incompleteness}. The completeness limit is defined as
the magnitude at which the narrow band data is 50\% complete and is
estimated by fitting the function

\begin{equation}
f(m) = \left({\rm exp}\left(\frac{m-m_c}{m_d}\right)+1\right)^{-1}
\label{eq:completeness}
\end{equation}

to the completeness fraction as a function of magnitude. The
parameters $m_c$ and $m_d$ are the 50\% completeness limit and a
measure of how quickly incompleteness sets in, respectively.  The 50\%
completeness limits of the survey fields are listed in Table
\ref{tab_gen}. For an object with magnitude $m_c$, the signal-to-noise
ratio over an 8 pixel diameter aperture is approximately 6. In
computing this signal-to-noise ratio, the correlation in the noise
between pixels has been taken into account. By creating an artifical image
that has the same pixel-to-pixel noise as the data, one finds no false
detections to the same detection threshold.

Objects that are detected only in the narrow band image are either
line emitting objects with large equivalent widths, extremely red
objects or false detections. In Fig.~\ref{fig:incompleteness}, these
objects are plotted in dot-dashed line. Their number does not become
significant until one passes the 50\% completeness limit. Beyond this
limit, most objects that are only seen in the narrow band image are
probably false. So we choose this limit as the limit of our survey.
With the exception of field F15, which is significantly shallower than
the other fields, the average limiting magnitude is 23.3,
corresponding to a flux (resp. luminosity) limit of $\sim 1.3 \times
10^{-17}$\,ergs\,s$^{-1}$\,cm$^{-2}$ (resp. $\sim 1.3 \times
10^{43}$\,ergs\,s$^{-1}$). The flux and luminosity limits do
not account for partial line transmission effects.

\subsection{Candidate selection}

Lyman alpha emitting candidates are selected using the following
criteria. They should be

\begin{itemize}

\item visible in narrow band ISAAC image and

\item not visible in the optical (F435W, F606W, F775W, and F850LP) ACS images

\end{itemize}

We note that these criteria do not make use of the IR data: no criterion was
set a priori for the equivalent width of Ly$\alpha$ at high redshift, and the
IR data were used a posteriori to remove low redshift objects (see
below). This step led to an initial list of ten candidates which we then
visually inspected in
all bands, including the ISAAC J, H and K band data. Down to the
limiting magnitudes listed in Table \ref{tab_gen}, all candidates were
found to be either

\begin{itemize}

\item ghosts related to bright stars (two objects)

\item in areas of low signal-to-noise that are near the edges of 
the combined narrow band images and are, therefore, outside our survey
area (one object), or

\item very red objects not detected in the optical bands with significant
  flux in the J, H and K ISAAC bands (five objects)


\end{itemize}


In the later category can be found Extremely Red Objects (EROs) (R - K
$>$ 5) or T-dwarfs which exhibit even redder colors (I-J $>$ 6). 
These objects can therefore pass our
initial selection criterion if they are detected in the NB image. EROs have
increasingly red colors from NB to K (J - K $>$ 1.8)
\citep{Cimatti2003}, whereas T-dwarfs have approximately flat J, H and
K colors and NB (1.19$\mu$m) - J $\sim$ 1 due to a strong water
absorption band at 1.19 $\mu$m. We then inspected all initial candidates and
further eliminated those which had broad band IR colors consistent with one or
the other of these very red objects. In practice, only objects consistent with
ERO colors were found.
Note that the data points of the dot-dashed line of figure 1 correspond to the
candidates {\it before} this second step was performed.

After this careful refinement of the candidate list, we were left with
no objects that were brighter than the 50\% detection limit, and two
candidates fainter than this limit, both only visible in the
NB1190 image. In view of the limiting magnitudes of the J and NB
images, this corresponds to $J - NB1190 \ga$ 1.5, which in turn
corresponds to objects with observed (resp. rest-frame) equivalent
widths $EW_{\rm obs} \ga 300\ \AA$ (resp. $EW_{\rm rest} \ga 30\ \AA$)
which are acceptably large values for a very high redshift starburst
galaxy.

Even though the magnitude of the two candidate were fainter
than the 50\% completeness limit, we decided to observe the most promising one
(NB1190=23.9 and S/N=3 in the F21 field) with
SINFONI with the aim of verifying that the object was real.

\section{SINFONI observations}

We used SINFONI on Yepun (VLT-UT4) on the night of 2005 October 31 to
observe the most promising candidate in the survey. SINFONI is a near
infrared integral field spectrograph \citep{Eisenhauer2003} that can
be fed with an adaptive optics (AO) corrected beam. Since there were no
bright stars near to the candidate, we used SINFONI without AO. We
used the 8\arcsec\ field-of-view, resulting in a pixel scale of
0\farcs125 perpendicular to the image slicer. The image slicer splits
the field-of-view into 32 slices, so each slice is 0\farcs25
seconds wide.  The 32 slices were dispersed with the J-band grating
and imaged onto a Hawaii 2k$\times$2k detector. The resulting spectra start
at 10950\,\AA\ and effectively end at the beginning of the strong telluric
absorption feature that starts at $\sim$ 13500\,\AA. The resolution of
the spectra is about 2000.

Individual exposures lasted 900 seconds, which is long enough to be
detector dark noise limited at 1.19 microns. A single observing
sequence consists of four such exposures with the object dithered
within the 8\arcsec\ field-of-view. At the end of the sequence, a
nearby star, which was used to verify the pointing, was observed. This
sequence was repeated 6 times and results in a total integration time
of 21600s.

The observing and reduction strategies were designed to maximise the
signal-to-noise ratio of a point-like emission line object. The first
step in the reduction process was to create a 2 dimensional sky frame
for each exposure. For any given exposure, the corresponding sky frame
was made from the average of the 6 other frames. Thus, instead of
reducing the optimal signal-to-noise ratio by a factor $\sqrt{2}$, which
would be the case if the sky subtraction was done on the basis of
image pairs, the signal-to-noise ratio is reduced by $\sim 8$\%. The maximum
number of frames that one can use for a sky is limited by instrument
flexure. We found that 6 frames, taken over a period of two hours, was
the best compromise between maximising the signal-to-noise ratio and
minimising the residuals caused by flexure.

The sky-subtracted data were then flat fielded and sectioned into 32
images, one for each slice. The 2 dimensional sky subtraction is not
perfect because, as a function of wavelength, the relative intensity
of the IR night sky varies with time. Night sky residuals were removed
by fitting and subtracting low order polynomials along the spatial
direction of the slices.

The resulting 2 dimensional sky-subtracted data were then registered
and averaged to produce a 2 dimensional co-added frame. A one square arc
second region was then extracted from the co-added frame to produce a 1
dimensional spectrum that was then wavelength and flux calibrated.

After six hours of integration, no emission line was detected. Our 5
sigma detection limit over a one square arc-second aperture is 
$ 1.5 \times 10^{-18}\,{\rm erg\,s^{-1}\,cm^{-2}\,pixel^{-1}}$, or 
$3.0\times 10^{-18}\,{\rm erg\,s^{-1}\,cm^{-2}}$ if we integrate over 6
Angstroms (4 pixels). From the 
magnitude of the candidate, we expected a flux of $7.0 \times
10^{-18}\,{\rm erg\,s^{-1}\,cm^{-2}}$. We conclude that our best candidate was
either spurious or a transient object (see next section).

\section{Contamination from transient objects}

Since the narrow band and broad band data were taken at different
times, it is possible for transient objects to appear in one
filter and not another. Our candidate selection is based on finding
sources that can only be seen in the ISAAC narrow band images.
Therefore, transient objects that are brighter than the narrow band
detection limit will be considered as candidates if they are not
visible in the broad band images. We consider two transients that
might contaminate the survey: supernovae and bodies in the solar system.

A few of the fields were observed at a time when the angular motion of
distant solar system bodies as seen from the Earth can be very low. In
individual exposures, which typically last 20 minutes, such objects
may appear point-like. Since the total exposure times were long and since the
observations were done in service, the data were taken during
different nights.  For field 21, the time over which the data were
taken spans 4 days; for field 16 it took 6 weeks to collect the data.
This, together with the location of the fields, which are 45 degrees
from the ecliptic, means that it is very unlikely that we would
mistakenly identify a distant body in the solar system as a high
redshift galaxy.  When the data are combined, objects in the solar
system will either be rejected when clipping for cosmic rays or appear
as trailed irregular-looking sources. In either case, it is
sufficient to split the data into subsets that are based on the date
of the exposure to test if a candidate is moving or stationary. We
have done this for all our initial candidates and no moving targets
were identified.

At the flux limit of the survey, distant supernovae can be visible for
several weeks, so supernovae are a potential source of contamination.
We computed the expected number of Type Ia and Type II supernovae that
would be visible in our narrow band ISAAC images by using the rates
published in \citet{Hopkins2006}. Even though Type II supernova
are approximately 5 times more numerous than Type Ia supernova, they
are, on average, 2.5 magnitudes fainter \citep{Filippenko1997}, so Type Ia
supernovae dominate the expected counts. Using a magnitude limit of
23.5 in the narrow band filter, we would need approximately 100 square
arc minutes to detect one Type Ia supernova. This is only three times the
area covered in our survey, so it is possible that there are Type Ia
supernova that are visible to the limit of the survey. However,
Type Ia supernova are usually close to their host galaxies. Since most
of these hosts would be detected in our broad band images, most Type
Ia supernovae would not be selected as candidates. There are, however,
a number of supernova that are considered as hostless. I.e. to the
limit at which imaging data is available, no host is visible.  From
Table 4 in \citet{Lidman2005} one estimates that 1 in 10 high
redshift Type Ia supernova can be considered as hostless. However, the
GOODS ACS imaging data is considerably deeper than what is
usually taken when following high redshift supernovae, so the fraction
of hostless Type Ia supernova is likely to be even less than 1 in 10.

Although it is very unlikely that we would have selected a supernova
of any type as a candidate in our survey, we cannot completely rule it
out. In surveys that cover larger areas or use less extensive broad
band imaging, contamination from transients such as supernova or
more unusual events, such as the one recently discovered in Bootes
\citep{Dawson2006}, might occur, so strategies to avoid this
contamination, such as taking the narrow band data over several epochs
separated by several months, should be considered.

\section{The $z \sim 8$ LAE luminosity function}
No Ly$\alpha$ emitters at z $\sim$ 8.8 have been found in 31 arcmin$^2$
down to $\sim 1.3 \times 10^{-17}$\,ergs\,s$^{-1}$\,cm$^{-2}$. Our results
complement the work of \citet{Willis2005} who performed similar
observations with the same filter and instrument, however with
different survey area and depth.

From a large sample of 58 Ly$\alpha$ emitters (LAEs) including 17
spectroscopic confirmations, \citet{Kashikawa2006} have established
the luminosity function of z = 6.5 LAEs. Similarly, \citet{Shimasaku2006} have
established the luminosity function (LF) of z = 5.7 Ly$\alpha$
emitters from a sample of 89 LAEs, of which 28 spectroscopically
confirmed. Unlike previous work \citep{Malhotra2004},
\citet{Kashikawa2006} show convincing evidence for evolution of the LF
between z = 5.7 and z = 6.5. It is not clear whether this
evolution can be attributed to the luminosity $L^*$, the comoving
number density $\Phi^*$, or the faint end slope $\alpha$, or a combination of
those, although it is suggested that $L^*$ is dimmed by a factor $\sim$ 2
between these two redshifts. 

\begin{figure}
\centerline{\epsfig{file=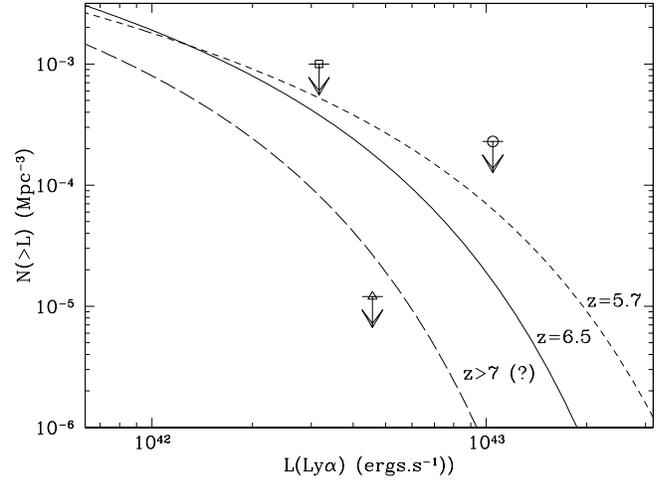,angle=-90,width=9.cm}}

\caption{Luminosity function of LAEs at z = 5.7 (dashed line,
\citet{Shimasaku2006}) and z = 6.5. (plain line, \citet{Kashikawa2006}).
Square: \citet{Willis2005}, Circle: this work, Triangle:
on-going survey at CFHT targeting z = 7.7 LAEs. Data points have not been
corrected for completeness. Extrapolation of the LF to higher z (z $>$ 7),
assuming the same evolution from z=6.5 as between z=5.7 and z=6.5 is shown for
illustration.} 
\label{fig:lf}
\end{figure}

In figure~\ref{fig:lf} we represent the LAE luminosity functions at
redshifts 5.7, 6.5 and at z $>$ 7 assuming arbitrarily another dimming
in $L^*$ by a factor 2 between redshifts 6.5 and $>$ 7. Also represented 
on this figure are the data points corresponding to this work
and to \citet{Willis2005}\footnote{We note that all data used in
  figure~\ref{fig:lf} suffer from very large error 
bars, not shown. One source of error comes from the way the comoving volume
sampled by the NB imaging survey is computed.
\citet{Willis2005} correctly compute the comoving volume as a function
of the Ly$\alpha$ luminosity (and line width), while other authors compute a
fixed comoving volume from the full width at half maximum of the NB
filter. Such differences could significantly affect the luminosity function
derived from number counts, and in particular the faint end slope of the LF. 
For consistency with other work, we have used the
traditional way of computing the comoving volume when plotting our datapoint
on figure~\ref{fig:lf}, and we adapted the \citet{Willis2005} datapoint
accordingly.}. The two data points are consistent with the
likely evolution of the luminosity function towards higher redshifts,
but do not allow us to constrain this evolution due to the limited
number of fields covered.  We also indicate on figure~\ref{fig:lf} a
point corresponding to an on-going programme that we are carrying out
at CFHT with a wide field imager (WIRCAM) at redshift 7.7.  When this
programme is complete, we should either detect a few z = 7.7 galaxies
if the luminosity function has moderately evolved from z = 6.5, or, in
case of no detections, infer a significant evolution of the LF.

Other wide field imagers
equipped with low OH NB filters, either at 1.06 $\mu$m or 1.19
$\mu$m, are becoming available on 4m or 8-10m class telescopes.
In 2007, the VLT will be equipped with
HAWK-I (see \citet{Casali2006} and http://www.eso.org/instruments/hawki), a $\sim$ 8\arcmin $\times$
8\arcmin\ imager, and VISTA (see \citet{McPherson2006} and 
http://www.vista.ac.uk), the new 4m IR
telescope to be installed at Paranal will cover a field of view of $\sim$
45\arcmin $\times$ 45\arcmin. DAZLE \citep{Horton2004}, a visitor instrument
dedicated to 
searching high-z LAEs (7\arcmin $\times$ 7\arcmin\ field of view equipped 
with very narrow band filters at $<$ 0.1\% bandpass) will observe at the VLT
in the coming months.

\begin{figure}
\centerline{\epsfig{file=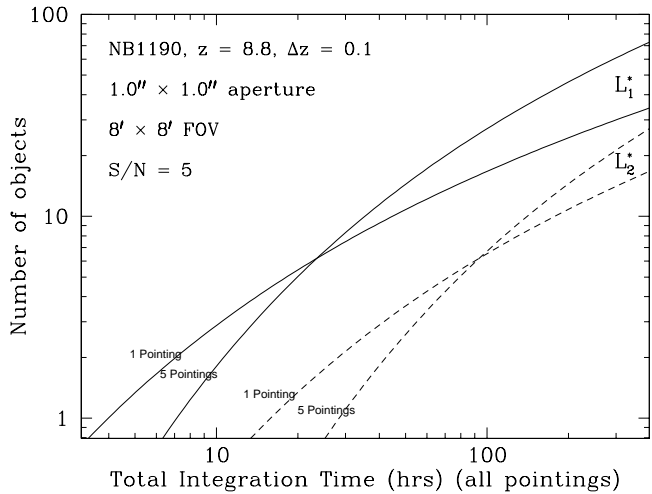,angle=-90,width=9.cm}}

\caption{Number of detected LAEs versus time. The VLT/HAWK-I case is
  assumed. Two different strategies (one deep single pointing and 5 shallower
  pointings for the same total time on sky) are indicated, for two luminosity
  functions: the z = 6.6 LF \citep{Kashikawa2006} (solid line, $\rm L_1^*$)
  and the same z $>$ 7 LF as in figure 2 
with $\rm L_1^*$ dimmed by a factor two (dashed lines,  $\rm  L_2^*$).}
\label{fig:survey}
\end{figure}

These new instruments will be essential for investigating high-z LAEs by
probing much larger volumes at similar or fainter detection limits. 
Different survey strategies can be contemplated with these instruments: 
either go very deep in a single field, or
alternatively go brighter in several fields
(pointings). Figure~\ref{fig:survey} shows the number of objects that can be
detected using two different survey strategies with HAWK-I at the VLT. 
Two luminosity functions are assumed: the z = 6.5 LF ($\rm L_1^*$)
\citep{Kashikawa2006} and the ad'hoc z $>$ 7 LF mentioned above ($\rm L_2^*$). 
The two different strategies, for each of the 
two luminosity functions, compare the number of objects detected in a single
`deep' field and in 5 non-overlapping `shallow' fields observed each for a 5
times shorter integration time.

The plot indicates that the `deep' strategy is naturally to be preferred for
short integration times, while favoring area over depth becomes
advantageous for very long integration times. Considering the very high
uncertainties  in the LF of z $>$ 7 galaxies, a single pointing
appears preferable for NB imaging surveys up to 100 hrs of integration time or
less for the instrumental characteristics considered in this example.

\section{Conclusion}
We have searched for z $\sim$ 8.8 Ly$\alpha$ emitters with a NB filter
centered at 1.19 $\mu$m in ISAAC at the ESO VLT. We covered an area of
31 arcmin$^2$ down to a line flux of $\sim 1.3 \times
10^{-17}$\,ergs\,s$^{-1}$\,cm$^{-2}$. No LAE was detected in this
pilot survey, consistent with recent estimates of the luminosity function at
z $\sim$ 6. Similar observations with wider field instruments are
likely to rapidly unveil high-z LAEs at redshifts above 7, however
in limited numbers and at high cost in observing time.
Systematic surveys of large numbers of very high z
galaxies will have to await JWST and ELTs.

\end{document}